\documentclass[preprint, aps, longbibliography]{revtex4-1}
\usepackage{amsmath}
\usepackage{graphicx}
\usepackage{layouts}
\usepackage{lineno}
\usepackage{amsmath}
\usepackage{amssymb}
\usepackage{subfigure}

\begin{document}
	
\title{The role of bulk eddy-viscosity variation on the log-layer mismatch observed in wall-modeled large-eddy simulations}

\author{Rey DeLeon}
\affiliation{Department of Mechanical Engineering \\University of Idaho, Moscow, Idaho 83844}
\author{Inanc Senocak}
\affiliation{Department of Mechanical Engineering and Materials Science \\University of Pittsburgh, Pittsburgh, PA 15261}

\date{December 20, 2017}

\begin{abstract}
	We investigate the role of the bulk eddy-viscosity variation on the well-known log-layer mismatch problem. An analysis of the mean momentum-balance  shows that the modeled stress term close to the wall can dominate because of the bulk eddy-viscosity. Consequently, the momentum-balance equation lacks a degree-of-freedom and the mean velocity conforms to an incorrect profile to satisfy the momentum-balance. We show that zonal enforcement of the target mass flow-rate can be an effective strategy to introduce an additional degree of freedom to the mean momentum-balance, which led to a significant reduction in the log-layer mismatch. When the mass flow-rate is enforced zonally, the filtered velocity field attains its own constant velocity-scale above the Reynolds-averaged field, supporting the hypothesis that there exists an artificial boundary layer above the Reynolds-averaged region. We simulate turbulent channel flows at friction Reynolds numbers of 2000 and 5200 on coarse meshes that would put the first point away from the wall well into the logarithmic layer. Second-order turbulence statistics and one-dimensional velocity spectra agree well with the direct numerical simulation benchmark data when results are normalized by the velocity-scale extracted from the filtered velocity field. Additionally, the error in the skin-friction coefficient for friction Reynolds numbers of 2000 decreased from $14.1\%$ to $2.5\%$ when we enforced the mass flow-rate zonally.
\end{abstract}

\maketitle

\section{Introduction}
The computational cost of large-eddy simulation (LES) of wall-bounded flows at high Reynolds numbers is expected to remain a challenge in the foreseeable future. The stringent resolution requirements arise from the scales of turbulence diminishing drastically as the wall is approached \cite{Pope2000}.  The computational burden of resolving the near-wall region can be relaxed through \textit{wall-modeled LES} (WMLES) \cite{Deardorff1970, Schumann1975}. When the viscous sublayer is not resolved, a shear stress boundary condition can replace the no-slip condition at the wall.  The shear stress at the wall is derived from flow variables with the help of the law-of-the-wall \cite{Schumann1975}.

One of the popular approaches to alleviate computational costs of LES is \textit{detached eddy simulation} (DES), first proposed by \citet {spalart1997} which was later refined in \citet{spalart2009detached} for high-Reynolds-number, massively-separated external aerodynamics where governing equations in the thin boundary layer are solved in the Reynolds-averaged Navier-Stokes (RANS) sense and LES is used away from the boundary layer where the grid size is adequate for resolving turbulent motions. \citet{nikitin2000} applied DES as a sub-grid scale model in an LES of turbulent channel flow. Turbulence was sustained but there was the appearance of a ``DES buffer layer'' where too steep velocity gradients created a drastic mean profile mismatch relative to the expected logarithmic velocity profile. The large velocity gradients were deemed to be the result of diminishing eddy viscosity when transitioning from the RANS near-wall flow to LES outer flow. Nikitin et al. observed that turbulent eddies did not form during this transition.  Large ``super-streaks,'' or eddies that are unrealistically large, have been observed in the transition region \cite{piomelli2003inner}. This issue of an artificial overestimation of the mean velocity profile in the LES region is later coined as the \textit{log-layer mismatch} problem, which is not specific to the DES model and has been reported in other wall-modeled LES studies as well.

\citet{shur2008hybrid} proposed an idea to combine the delayed DES (DDES) approach of  \citet{spalart2006new} for areas where grid resolution prevents resolving dominant eddies and WMLES for grid resolutions that can resolve dominant eddies.  The method, referred to as improved DDES (IDDES), blends together these two approaches. IDDES performed just as well in massively separated flows as DES or DDES (the latter improved upon DES by allowing for more grid refinement and thicker boundary layers), and improved upon the log-layer mismatch in plane channel flow. IDDES also performed very well in flows with both attached and separated flow regions.

\citet{renard2015improvements} used zonal DES (ZDES) \cite{deck2012recent} for WMLES of a zero-pressure-gradient turbulent boundary layer with an improved positioning for the RANS-LES interface by placing the transition in the geometric center of the logarithmic layer. Good agreement was achieved with skin friction correlations. ZDES differs from DDES or IDDES in that the user determines the regions of RANS and LES as opposed to a length scale dependent on grid resolution. As is a common challenge with user-defined zonal approaches, determining the RANS-LES interface placement will become more involved with more complex geometries.

\citet{baggett1998feasibility} performed numerical tests with a turbulent channel flow case to study the artificial turbulent layer that forms between a Reynolds-averaged near-wall region and a filtered velocity field above that region. Baggett argued that a Reynolds-averaged field is effectively laminar and therefore cannot provide the turbulent fluctuations to the filtered velocity field. To overcome this shortcomings, both \citet{keating2006dynamic} and \citet{Davidson2006} introduced a stochastic-type external forcing at the interface of Reynolds-averaged and filtered velocity fields to generate fluctuations that mimick Reynolds stresses.  \citet{keating2006dynamic} used a proportional controller to force the flow based on the difference between resolved and modeled Reynolds stress in the transition zone. This approach generates enough Reynolds stress to correct the log layer mismatch and break up the so-called ``super-streaks'' that occur in transition regions \cite{piomelli2003inner}. \citet{radhakrishnan2006reynolds} later observed accurate results without stochastic forcing in unstable flows where the mean values tend to have perturbations. \citet{Davidson2006} introduced turbulent fluctuations that were obtained from synthesized homogeneous turbulence.   
 
Most recently, \citet{yang_etal_2017} revisited the log-layer mismatch problem and proposed a local temporal filtering when calculating the wall shear stress. They argued that the temporal filtering breaks the unphysical correlation between wall shear stress fluctuations and the velocity fluctuations in WMLES. Their remedy reduced the log-layer mismatch. However, the authors admitted that wall-shear stress fluctuations did not improve, and they did not present any profiles of second-order statistics of turbulence across the channel height. Yang et al. alluded to the potential role of bulk eddy-viscosity on the log-layer mismatch, but they did not elaborate on it when explaining their remedy. This is most likely because they use the scale-dependent Lagrangian dynamic Smagorinsky model \cite{scale_dependent} that does not create a significant variation in the bulk eddy-viscosity. In the current study, we follow the same baseline formulation as Yang et al., but use the scale-independent Lagrangian dynamic Smagorinsky model \cite{Meneveau1996} and introduce the bulk eddy-viscosity variation into the formulation by switching to a mixing-length model in close proximity of the wall. We enforce the target mass flow-rate zonally, which partially decouples the filtered velocity field from the Reynolds-averaged field. As an outcome of this partially decoupled approach, the log-layer mismatch reduces substantially. Furthermore, the filtered velocity region attains its own velocity-scale, supporting the argument of \citet{baggett1998feasibility} that an artificial boundary layer forms when the filtered velocity field is allowed to develop over a Reynolds-averaged field. Second-order statistics of turbulence and velocity spectra show improvements when normalized by this velocity-scale extracted from the filtered velocity region.
\section{Numerical Formulation}

We solve the following filtered form of the governing equations for incompressible flows.
\begin{align}
\label{eq:Continuity}
\frac{ \partial \overline{u}_j }{ \partial x_j } = 0,\\
\frac{\partial \overline{u}_i}{\partial t} + \frac{\partial}{\partial x_j} \left( \overline{u}_i \overline{u}_j \right) =  -\frac{1}{\rho} \frac{\partial \overline{p}}{\partial x_i} & + \frac{\partial}{\partial x_j} \left( 2 \nu \overline{S}_{ij} - \tau _{ij} \right) + f_i,
\label{eq:MomEqns}
\end{align}
where
\begin{align}
\label{eq:StressTensor}
S_{ij} = \frac{1}{2} \left( \frac{ \partial \overline{u}_i }{ \partial x_j } + \frac{ \partial \overline{u}_j }{ \partial x_i } \right)
\end{align}
is the strain rate tensor, and  
\begin{align}
\label{eq:SGStensor}
\tau _{ij} = \overline{u_i u_j} - \overline{u}_i \overline{u}_j
\end{align}
is the tensor for the SGS Reynold stresses, and $f_i$ is a body force term.

\citet{yang_etal_2017} used a scale-dependent Lagrangian dynamic Smagorinsky model with the wall shear stress boundary condition determined from a logarithmic-law \cite{Schumann1975,Grotzbach1987}. Here, we follow the same baseline formulation, except that we adopt the scale-independent Lagrangian dynamic Smagorinsky eddy viscosity model \cite{Meneveau1996} for LES. In the Reynolds-averaged region, we use the mixing length model of \citet {Prandtl1925} as it is an adequate model for attached flows. To introduce a significant bulk eddy-viscosity variation in the momentum equations, we combine the LES filter-size with the mixing-length scale as follows:
\begin{align}
\label{eq:exponential}
l_{mix} = \left[ 1 - \exp \left( -y / h_{RL} \right) \right] & C_S \Delta + \exp \left( -y / h_{RL} \right) \kappa y, \\
\label{eq:blendedEddyViscosity}
\nu_t = & ~l_{mix}^2 \vert \overline{S} \vert, \\
\tau _{ij} = &  \nu_t S_{ij},
\end{align}
where $h_{RL}$ is the height that controls the extend of the Reynolds-averaged region in the wall-normal direction. The current model is not intended to simulate turbulent flows over complex geometry, but it is sufficient to investigate the role of the bulk eddy-viscosity on the log-layer mismatch problem in a turbulent channel flow.

We aim to place the first off-wall grid point in the logarithmic region ($y^+ >$ 30) to study wall-modeled LES on coarse meshes. Direct application of the no-slip boundary condition on such coarse meshes would not be adequate. Instead, we impose a shear-stress boundary condition using the wall-model of Schumann\cite{Schumann1975}, which was later modified to a more general form by Gr{\"o}tzbach\cite{Grotzbach1987}. The essence of the Schumann-Gr{\"o}tzbach model is to provide an instantaneous shear-stress boundary condition at the wall by providing components of the total wall shear stress as follows:

\begin{align}
\overline{\tau}_{xy, w} \left( x,z,t \right) & = \frac{ \overline{ u } \left( x,y_1,z,t \right) }{ \langle \overline{u} \left( y_1 \right) \rangle } \langle \overline{\tau}_{xy,w} \rangle, \\
\overline{v} & = 0, \\
\overline{\tau}_{zy, w} \left( x,z,t \right) & = \frac{ \overline{ w } \left( x,y_1,z,t \right) }{ \langle \overline{u} \left( y_1 \right) \rangle } \langle \overline{\tau}_{xy,w} \rangle,
\end{align}
where $y_1$ is the first off-wall grid point.  Schumann originally obtained the mean wall shear stress from the momentum balance in a channel flow, $\langle \overline{\tau}_{xy,w} \rangle = - \delta \langle \frac{\partial \overline{p}}{\partial x} \rangle$, where $\langle \frac{\partial \overline{p}}{\partial x} \rangle$ was given \textit{a priori}. Gr\"{o}tzbach generalized this wall-model by determining the stress from the log-law:
\begin{align}
\label{eq:GrotzLogLaw}
\frac{ \langle \overline{u} \left( y_1 \right) \rangle }{ u_{\tau} } & = \frac{1}{\kappa} \ln \left( \frac{y_1 u_{\tau}}{\nu} \right) + B, \\
\label{eq:MeanWallShearStress}
\langle \overline{\tau}_{xy,w} \rangle & = \rho u_{\tau}^2,
\end{align}
where $\kappa$ is the von K\'{a}rm\'{a}n constant, $B$ is a constant, and $\rho$ is the density, set to unity in this study.  While eliminating the need for an \textit{a priori} $\langle \frac{\partial \overline{p}}{\partial x} \rangle$, Gr\"{o}tzbach's modification requires knowledge of $\kappa$ and $B$. For this study, we adopt the commonly used values of $\kappa$ = 0.41 and $B$ = 5.2, but various other possible values exist \cite{Pope2000, nagib2008variations, bailey2014estimating, lee2015direct}.

To avoid contamination from discretization and SGS modeling errors near the wall \cite{cabot2000approximate, nicoud2001large, kawai2012wall}, Eq. \ref{eq:GrotzLogLaw} is solved at the third off-wall grid point at every time step such that $y_1$ becomes $y_3$ \cite{LeeJungil2013}. The strain rate tensor in Eq. \ref{eq:blendedEddyViscosity} requires the computation of wall-normal derivatives and will be incorrect due to the coarseness of the grid if calculated explicitly.  Therefore, the eddy viscosity in the first off-wall grid cell is computed via linear extrapolation from the second and third off-wall grid cells \cite{cabot2000approximate}. 


The governing equations are solved using a GPU-accelerated, three dimensional incompressible flow solver \cite{Thibault12, jacobsen2013multi, deleon2013large}. The code uses the projection algorithm \cite{Chorin1968} on directionally-uniform Cartesian grids with second-order central differences for spatial derivatives and a second-order Adams-Bashforth scheme for time advancement. The pressure Poisson equation is solved by an amalgamated parallel 3D geometric multigrid solver designed for GPU clusters \cite{Jacobsen2011}.

\subsection{Zonal Enforcement of the Mass Flow-rate}

As periodic boundary conditions alone cannot sustain a constant mass flow-rate in the channel, a body-force is needed to sustain the flow in the channel. There are two approaches to achieve this: prescribe a constant body-force that would balance the target shear stress at the wall, or dynamically adjust a uniform body-force to maintain a target mass flow-rate through the channel. We, like numerous other studies, adopt the latter approach, which is preferable when the wall shear stress is not known \textit{a priori}. Turbulence statistics also converge faster with the latter approach. It is commonly adopted in turbulent channel flow simulations, including DNS studies \cite{lee2015direct}. The particular technique we use is that of Benocci and Pinelli \cite{Benocci1990}, which can be formulated as follows:
\begin{align}
\label{eq:mdotcorr}
f^{t+1}_x = f^t_x - \frac{2}{\Delta t} \left( \frac{ \dot{m}^{t+1} }{A_c} - \frac{ \dot{m}^0 }{A_c} \right) + \frac{1}{\Delta t} \left( \frac{ \dot{m}^t }{A_c} - \frac{ \dot{m}^0 }{A_c} \right),
\end{align}
where the superscript $t$ is the time-step level, $\Delta t$ is the physical time-step value (which needs to be held constant throughout the simulation as per the original formulation), $\dot{m}^0$ is the prescribed mass flow-rate, and $A_c$ is the constant cross-sectional area.  Other formulations to maintain constant mass flow-rate can be found in the literature \cite{you2000modified, Brandt2003}.


\begin{figure}
	\centering
	\subfigure[]{\includegraphics[width=0.48\textwidth]{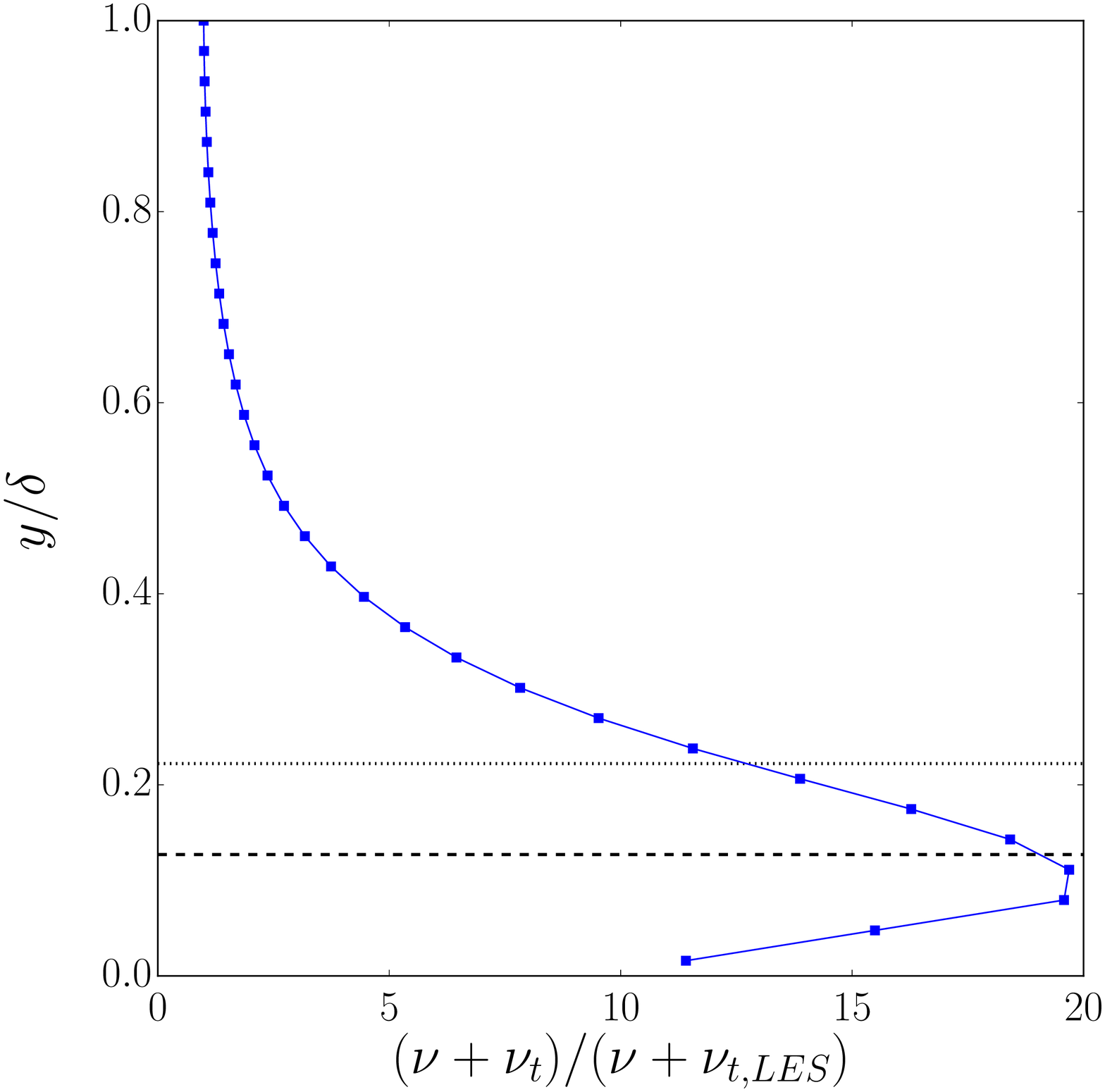}\label{fig:EddyViscTriForcing}}
	\subfigure[]{\includegraphics[width=0.48\textwidth]{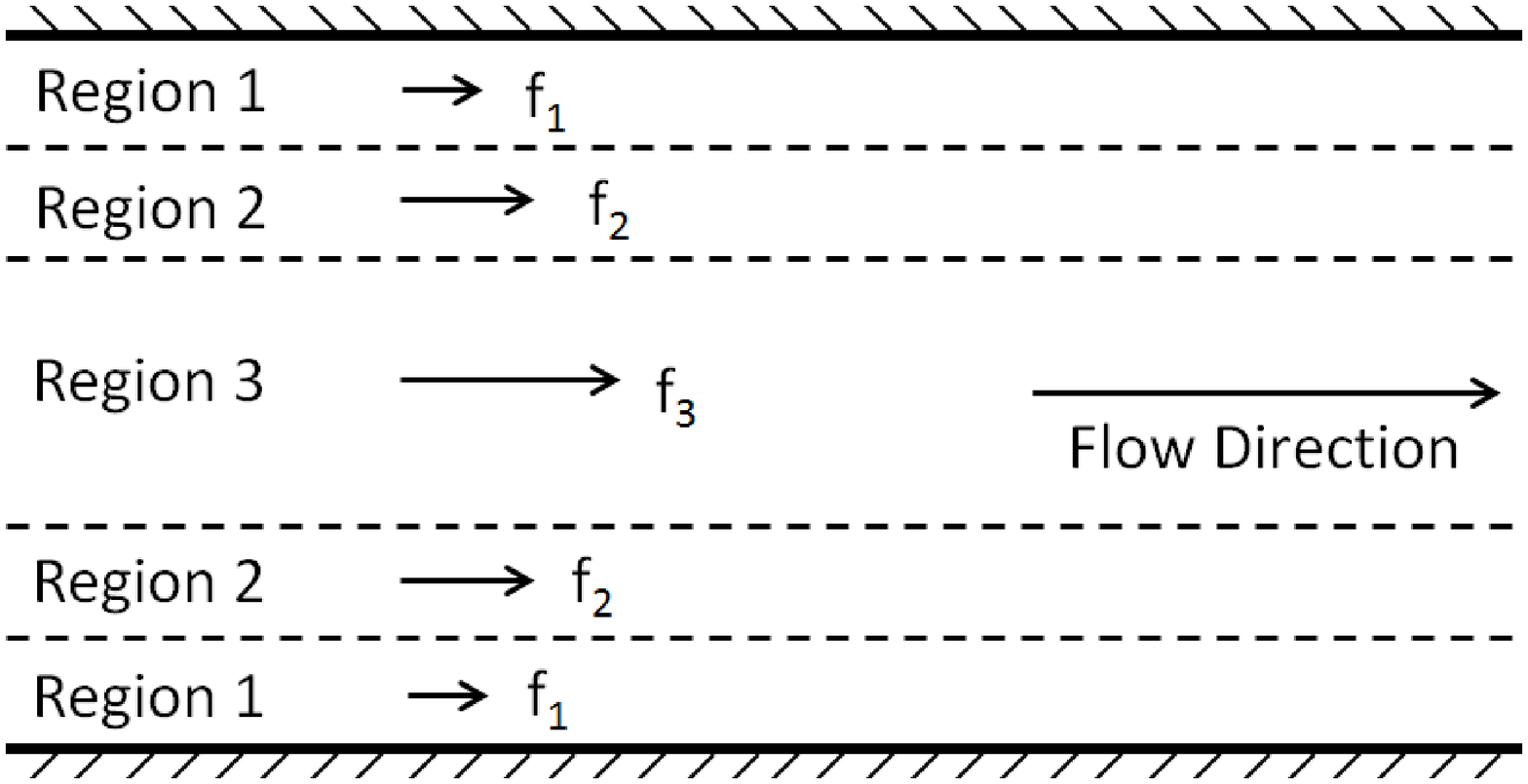}\label{fig:splitForcingSketch}}
	\caption{Split-forcing demarcations. (\textit{a}) The normalized mean eddy viscosity profile obtained from a RANS-LES simulation of Re$_{\tau}$ = 2000 turbulent channel flow adopting the eddy viscosity model given in Eq. \ref{eq:blendedEddyViscosity}. The dashed line represents the height of the RANS-LES interface, $h_{RL}$, and the first split-forcing demarcation. The dotted line represents the second split-forcing demarcation, $h_{sp}$. (\textit{b}) Sketch of three forcing regions in a channel flow.  }
\end{figure}

The log-layer mismatch observed in turbulent channel flow simulations with eddy-viscosity-based RANS-LES methods has been attributed to a lack of resolved Reynolds shear stress provided to the LES zone near the RANS-LES interface \cite{Davidson2006, keating2006dynamic, larsson2007artificial}. Here, we investigate the role of bulk eddy-viscosity variation on the log-layer mismatch and suggest a split-forcing approach grounded in the mass conservation principle to reduce it. 

For a fully-developed, statistically-stationary channel flow, the mean streamwise momentum equation reduces to the following
\begin{align}
\frac{d \tau}{dy} = -\frac{dp}{dx}.
\label{eq:channelMeanBalance}
\end{align}
The driving mean pressure gradient is constant on the right hand side of Eq. \ref{eq:channelMeanBalance} \cite{Pope2000}.  Integrating Eq. \ref{eq:channelMeanBalance} from the wall ($y$ = 0) to some wall-normal distance $y$ and solving for $\tau_w$, we obtain,

\begin{align}
\tau_w =
\underbrace{ \frac{dp}{dx}y                             \rule[-12pt]{0pt}{3pt}}_{\text{I}}
+\underbrace{\left. \nu \frac{dU}{dy} \right\vert_y              \rule[-12pt]{0pt}{3pt}}_{\text{II}}
+\underbrace{\left. \nu_t(y) \frac{dU}{dy} \right\vert_y        \rule[-12pt]{0pt}{3pt}}_{\text{III}}
-\underbrace{\left. \left\langle u v \right\rangle \right\vert_y \rule[-12pt]{0pt}{3pt}}_{\text{IV}}\,,
\label{eq:shearStressBalance}
\end{align}
where term I represents the driving pressure gradient, term II represents the viscous shear stress, term III is the shear stress modeled with an eddy-viscosity approach, and term IV represents the resolved Reynolds shear stress.  In a wall-resolved LES at a wall-normal distance sufficient to ignore viscous effects (i.e. ignore term II), we can argue that Eq. \ref{eq:shearStressBalance} has two degrees of freedom to achieve force balance through terms III and IV as $\frac{dp}{dx}$ is constant ($\frac{dp}{dx}$ is $f_x$ in Eq. \ref{eq:mdotcorr}).

We now evaluate Eq. \ref{eq:shearStressBalance} when the wall-normal variation of the bulk eddy-viscosity is significant, taking on a large value in the RANS region and a small value in the LES region. A typical eddy-viscosity profile that is relevant to the present analysis is shown in Fig. \ref{eq:blendedEddyViscosity}  We place the RANS-LES transition height sufficiently away from the wall to safely ignore term II. In theory, the RANS turbulence model simulates the ensemble-averaged field, thus approaching the interface from the RANS side means negligible resolved stresses, therefore term IV is negligible within the RANS region. As a result, the previous two degrees of freedom in a wall-resolved LES (terms III and IV) are now reduced to one (term III) when the near-wall region is modeled with a RANS model. 

The variation of the bulk eddy-viscosity, $\nu_{t}$, in the wall-normal direction in term III can be substantial as opposed to the sub-grid-scale (SGS) eddy-viscosity in a wall-resolved LES. Consequently, the mean velocity profile can assume unphysical profiles to satisfy the force balance within this single term. As numerous studies have shown, introducing an external forcing term around the RANS-LES interface reduces the log-layer mismatch markedly. The effectiveness of this approach becomes clear as this external forcing term introduces a second degree of freedom to the force balance. 

Loosely speaking, we can imagine a less viscous fluid (i.e. LES region) flowing over a highly viscous fluid (i.e. RANS region) when a RANS-LES model is adopted. Based on this analogy, we propose to impose the constant mass flow-rate through the channel zonally. Our hypothesis is that by imposing the mass flow-rate in party, we can introduce a second degree of freedom in Eq. \ref{eq:shearStressBalance}) without the need for a separate forcing term. The shear stress that arise between zones that are forced separately to maintain the target mass flow-rate would be sufficient to energize the scales that are damped by a large value of the eddy-viscosity. Our proposal for the channel flow is straightforward as one can derive the mass flow-rates in each zone from the logarithmic law-of-the-wall or benchmark data. However, extension to general flow configurations is admittedly not easy and beyond the scope of current investigation to understand the role of bulk eddy-viscosity variation on the log-layer mismatch.

We sketched the split-forcing approach as applied to a channel flow in Fig. \ref{fig:splitForcingSketch}. The regions are reflected about the centerline to enforce symmetry. Figure \ref{fig:splitForcingSketch} shows three streamwise forcing regions, $f_1$, $f_2$ and $f_3$.  Each forcing region has its own target mass flow-rate such that the mass flow-rate for the entire channel remains unchanged. We adjust Eq. \ref{eq:mdotcorr} by introducing a subscript, $n$, as an index for the forcing region as follows:
\begin{align}
	f^{t+1}_{n} = f^t_{n} - \frac{2}{\Delta t} \left( \frac{ \dot{m}^{t+1}_n }{A_c} - \frac{ \dot{m}^0_n }{A_c} \right) + \frac{1}{\Delta t} \left( \frac{ \dot{m}^t_n }{A_c} - \frac{ \dot{m}^0_n }{A_c} \right),
	\label{eq:forcing}
\end{align}
where $\dot{m}^t_n$ now represents the mass flow-rate through a particular forcing region. We prescribe the target mass flow-rate for each forcing region following a target mean profile, usually obtained from benchmark data. Target mass flow-rate for the channel is ensured by the following constraint: 
\begin{align}
\dot{m}_T^0 = \sum_{n=1}^{N} \dot{m}_n^0,
\label{eq:SumMdot}
\end{align}
where $\dot{m}_T^0$ is the total mass flow-rate through the channel and $N$ is the total number of forcing regions.

By enforcing the mass flow-rate zonally under the constraint of the target mass flow-rate for the entire channel, we induce stochastic shear stresses between the zones and energize the scales of the motion without resorting to an explicit stochasting forcing term. The stochastic forcing, which is implicit in our zonal mass flow-rate approach, can be viewed as follows:
\begin{align}
\label{xxx}
f^{\xi}_{x,n} = f^{t+1}_{n} - {u^{2}_{\tau,w} \over \delta},
\end{align} 
where the second term on the right hand side is the mean pressure gradient to sustain a flow rate at a prescribed friction Reynolds number.

We found that the results are not very sensitive to the exact location of the forcing regions. It is sufficient to inspect the mean eddy viscosity profile using a spatially-uniform forcing, which we refer to as \textit{single-forcing}, and split the domain based on the magnitude of bulk eddy-viscosity. Fig. \ref{fig:EddyViscTriForcing} shows the mean eddy viscosity profile from a single-forcing RANS-LES for Re$_{\tau}$ = 2000 channel flow. The simulation parameters for this case are given in Table \ref{tab:dims}. The dashed line in Fig. \ref{fig:EddyViscTriForcing} is the RANS-LES interface. The mean eddy-viscosity is normalized by its value in the LES region, which has a nearly constant profile. Maximum value of eddy viscosity occurs near the RANS-LES interface, $h_{RL}$, and is a logical place for a split-forcing demarcation. We initially attempted using only two forcing regions but the results did not show satisfactory improvement because the eddy viscosity is still large above the RANS-LES interface relative to the LES value and has decreasing trend. A second splitting height, $h_{sp}$, is placed in the large eddy viscosity region approximately where the eddy viscosity is 10 to 12 times that of the core value.  The hypothesis is that one more additional forcing region will introduce enough shear stress in the blending region to provide satisfactory results. The use of three forcing regions is referred to as \textit{tri-split}. We use approximately the same splitting heights in nondimensional wall units for the Re$_{\tau}$ = 5200 case (see Tab. \ref{tab:dims}).  Note that using more forcing regions than three may not be necessary as the goal of this experiment is to force regions of high eddy viscosity independent of the LES core where the eddy viscosity profile is nearly constant. Also, when eddy viscosity is nearly uniform, splitting the mass flow-rate would run against the theory described by Eq. \ref{eq:shearStressBalance}.


\section{Results and Discussion}

\begin{table}
	\begin{center}
	\caption{Simulation parameters. The superscript, $^+$, indicates nondimensional quantities in wall units. $y_1^+$ is the distance of the first off-wall streamwise velocity component.} 
	\begin{tabular}{ c c c c c c c c c c c }
		\hline 
		Re$_{\tau}$ & N$_x$ & N$_y$ & N$_z$ & L$_x$ & L$_y$ & L$_z$ & $\Delta x^+$ & $\Delta y^+$ & $\Delta z^+$ & $ y_1^+$ \\ 
		\hline
		2000 & 257 & 65 & 193 & 8$\pi \delta$ & 2 $\delta$ & 3 $\pi \delta$ & 200 & 64 & 100 & 32 \\ 
		5200 & 513 & 129 & 513 & 8$\pi \delta$ & 2 $\delta$ & 3 $\pi \delta$ & 250 & 82 & 100 & 41 \\
		\hline 
	\end{tabular}
	\label{tab:dims}
	\end{center}
\end{table}

\begin{table}
	\begin{center}
	\caption{Split-forcing parameters. The superscript, $^+$, indicates nondimensional quantities in wall units. $h_{RL}$ is the RANS-LES transition height and $h_{sp}$ is the splitting height applicable to tri-split forcing.} 
	\begin{tabular}{ c c c c c }
		\hline 
		Re$_{\tau}$ & $h_{RL}^+$ & $h_{sp}^+$ & $h_{RL}$ / $\delta$  & $h_{sp} / \delta$\\ 
		\hline
		2000 & 250 & 440 & 0.127 & 0.222\\ 
		5200 & 250 & 490 & 0.047 & 0.094\\
		\hline 
	\end{tabular}
	\label{tab:splitdims}
	\end{center}
\end{table}

Tables \ref{tab:dims} and \ref{tab:splitdims} present the simulation parameters and the split-forcing parameters for each turbulent channel flow case, respectively. The friction Reynolds number is based on friction velocity, $u_{\tau}$, and channel half-height, $\delta$.  We compare Re$_{\tau}$ = 2000 channel flow to the readily-available DNS data of Hoyas and Jimenez\cite{Hoyas2006}, and Re$_{\tau}$ = 5200, to the recent DNS data of Lee and Moser\cite{lee2015direct}. In all of our simulations, periodic boundary conditions were applied in the streamwise and spanwise directions (x-direction and z-direction, respectively).  Spin-up time for turbulence was 200 $\delta$/$u_{\tau}$ and sampling for turbulence statistics took place over an additional 200 $\delta$/$u_{\tau}$. 

\begin{figure}
	\centering
	\subfigure{\includegraphics[width=0.48\textwidth]{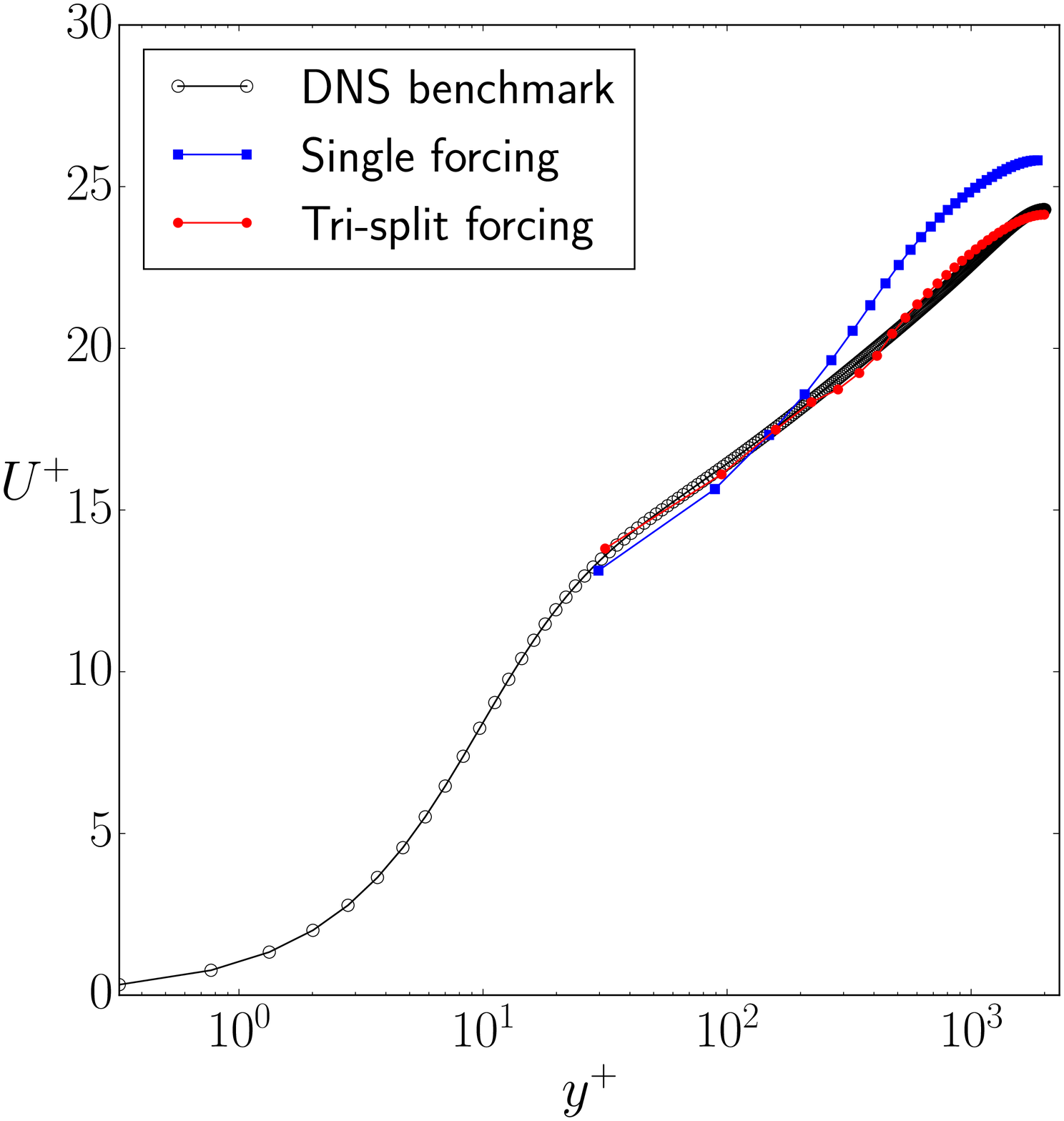}\label{fig:ForcingProgressionMean}}
	\subfigure{\includegraphics[width=0.48\textwidth]{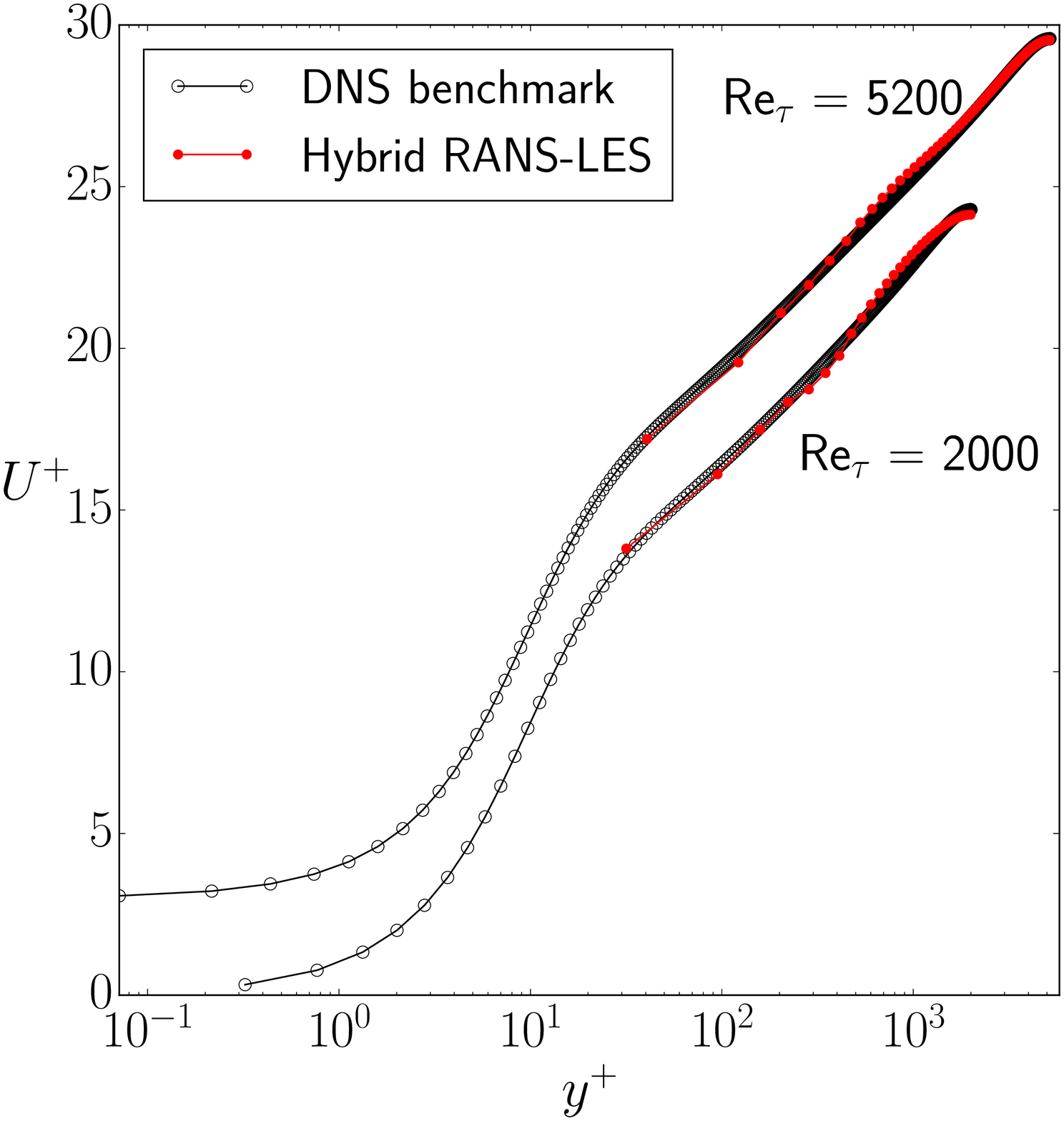}\label{fig:MeanVelocityWallNormalization}}
	\caption{Nondimensional mean velocity profiles. (\textit{a}) Log-layer mismatch reproduced using the single forcing approach. Re$_{\tau}$ = 2000. (\textit{b}) Re$_{\tau}$ = 5200 (shifted up by 3.0) and Re$_{\tau}$ = 2000 simulated with the tri-split approach. All results are normalized by $u_{\tau,w}$ obtained from the wall shear stress.}
\end{figure}

The effect of split-forcing on the mean velocity profile is shown in Fig. \ref{fig:ForcingProgressionMean} for  Re$_{\tau}$ = 2000. The single forcing of the channel flow reproduces the well-known log-layer mismatch problem. The tri-split forcing significantly reduces the mismatch from the DNS mean velocity, demonstrating a clear improvement over the single forcing approach. The effect of split-forcing is to introduce shear stresses into the flow to compensate for the lack of resolved shear stress in the RANS-LES interface region and enable the mean velocity field to match the expected profile.  In Fig. \ref{fig:MeanVelocityWallNormalization} we demonstrate the effectiveness of our tri-split forcing approach for an even higher Reynolds number flow, Re$_{\tau}$ = 5200. The corresponding bulk Reynolds number (based on channel height) for Re$_{\tau}$ = 5200 is approximately 250,000. Both profiles agree very well with the DNS benchmark data despite a mesh resolution that is coarse in the wall normal direction ($y_1^+ \approx 30$). For context, Lee and Moser \cite{lee2015direct} used roughly 121 billion mesh points for DNS of Re$_{\tau}$ = 5200 whereas we used 33.9 million mesh points without any stretching in the mesh. 

\begin{table}
	\caption{Comparison of computed skin friction coefficient with Dean's correlation\cite{dean1978reynolds}. Re$_b$ is bulk Reynolds number. Computed skin friction coefficient: $C_{f, comp} = \tau_w / \frac{1}{2}\rho U_b^2$. Dean's correlation for skin friction coefficient $C_{f,Dean} = 0.073 Re_b^{-0.25}$.}
	\centering
	\begin{tabular}{ c c c c c c c }
		\hline Re$_{\tau}$ & Forcing & U$_b$ / $u_{\tau,w}$ & Re$_b$ & $C_{f,comp}$  & $C_{f,Dean}$ & \% error \\ 
		\hline
		2000 & Single & 23.44 & 87793 & 3.64 $\times$ 10$^{-3}$ & 4.24 $\times$ 10$^{-3}$ & 14.1 \\
		2000 & Tri-split & 21.99 & 87698 & 4.14 $\times$ 10$^{-3}$ & 4.24 $\times$ 10$^{-3}$ & 2.5 \\
		5200 & Tri-split & 24.26 & 250543 & 3.40 $\times$ 10$^{-3}$ & 3.26 $\times$ 10$^{-3}$ & 4.1
	\end{tabular}
	\label{tab:deansCf}
\end{table}

Tri-split forcing improves skin friction estimates relative to the single forcing simulation. Table \ref{tab:deansCf} compares the skin friction coefficient, $C_{f}$, computed from the simulations and the Dean's correlation based on the bulk Reynolds number \cite{dean1978reynolds}.  The bulk velocity, $U_{b}$, is the averaged velocity through the channel cross section, and the bulk Reynolds number, $Re_{b}$, is based on the full channel height.  The percent error presumes the Dean correlation as the correct value. There is a notable reduction in percent error of tri-split forcing over single forcing with skin friction coefficient calculations in the Re$_{\tau}$ = 2000 case.  The Re$_{\tau}$ = 5200 also gives a small percent error. These values are much lower than the reported error values found in early studies \cite{nikitin2000}, which were approximately 10\% and above.

\begin{figure}
	\centering
	\subfigure{\includegraphics[width=0.48\textwidth]{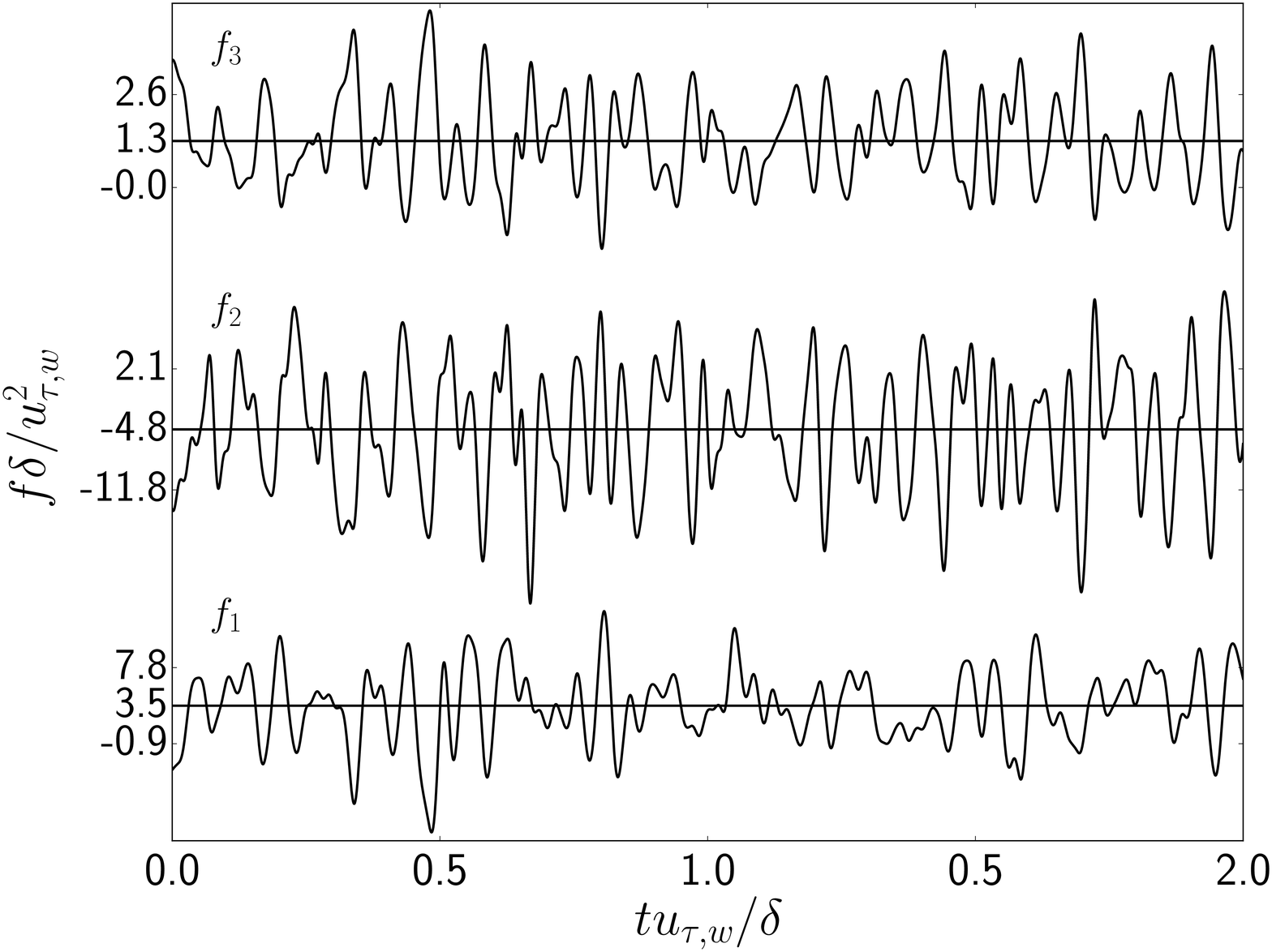}\label{fig:ForcingTimeSeries2000}}
	\subfigure{\includegraphics[width=0.48\textwidth]{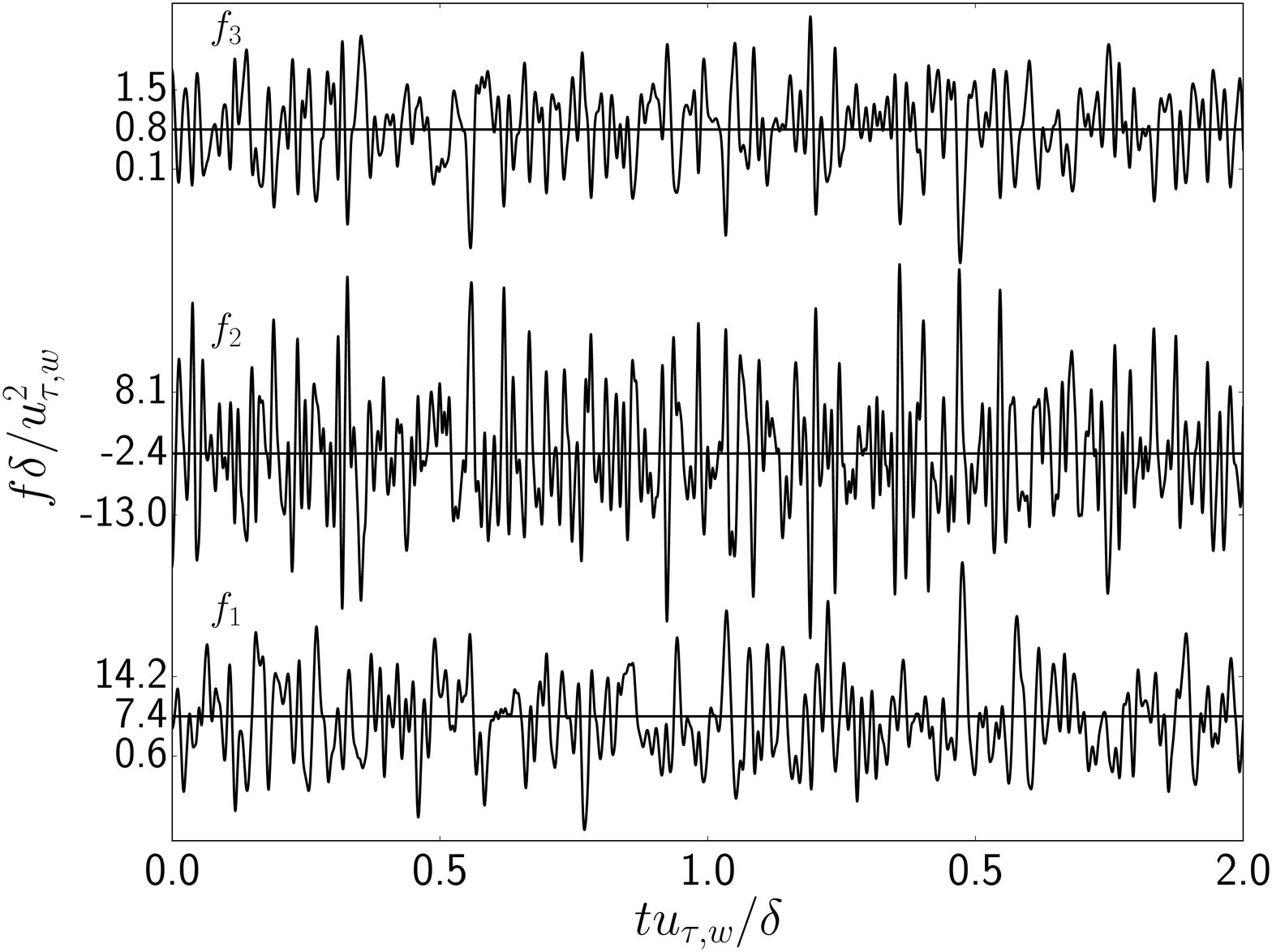}\label{fig:ForcingTimeSeries5200}}
	\caption{Tri-split forcing time series for (\textit{a}) Re$_{\tau}$ = 2000 and (\textit{b}) Re$_{\tau}$ = 5200 normalized by $\delta$ and $u_{\tau,w}$.  Results are shown over two eddy turnover times, $\delta$/$u_{\tau,w}$. The bottom time series in the figure is the near-wall region and the topmost time series is for the core forcing region.  The solid horizontal line is the mean of the time series found over the entire statistical sampling period.  The other two tickmarks are one standard deviation above and below the mean found over entire statistical sampling period.}
\end{figure}

Figures \ref{fig:ForcingTimeSeries2000} and \ref{fig:ForcingTimeSeries5200} show a time series of the forcing values for the tri-split for for both Re$_{\tau}$ = 2000 and Re$_{\tau}$ = 5200, respectively, over a time lapse of two eddy turnover times, $\delta$/$u_{\tau,w}$. We plot the evolution of the forcing in each zone to better understand the impact of split-forcing on the flow field. As will be discussed later, the wall shear stress can be recovered by converting the mean forcing values to a shear stress by integrating along channel height. The time series shows that the near-wall region and the middle forcing region have much greater variations than the core forcing region, with the middle forcing region being predominantly negative in value while the other two regions are positive in value. The magnitude of the eddy viscosity explains this behavior.  In the RANS region, the eddy viscosity is dominantly defined by the mixing length model, exhibiting a positive gradient.  The core forcing region, above $h_{sp}$, the eddy viscosity is dominantly defined by the sub-grid scale model and does not vary significantly.  However, the second forcing region between $h_{RL}$ and $h_{sp}$ has a negative gradient in eddy viscosity, as the model is decreasing from a RANS value to an LES value, leading to the middle forcing region being predominantly negative in value. The opposite signs of the forcing help energize the interface between each zone stochastically. 

\begin{figure}

\includegraphics[width=0.75\textwidth]{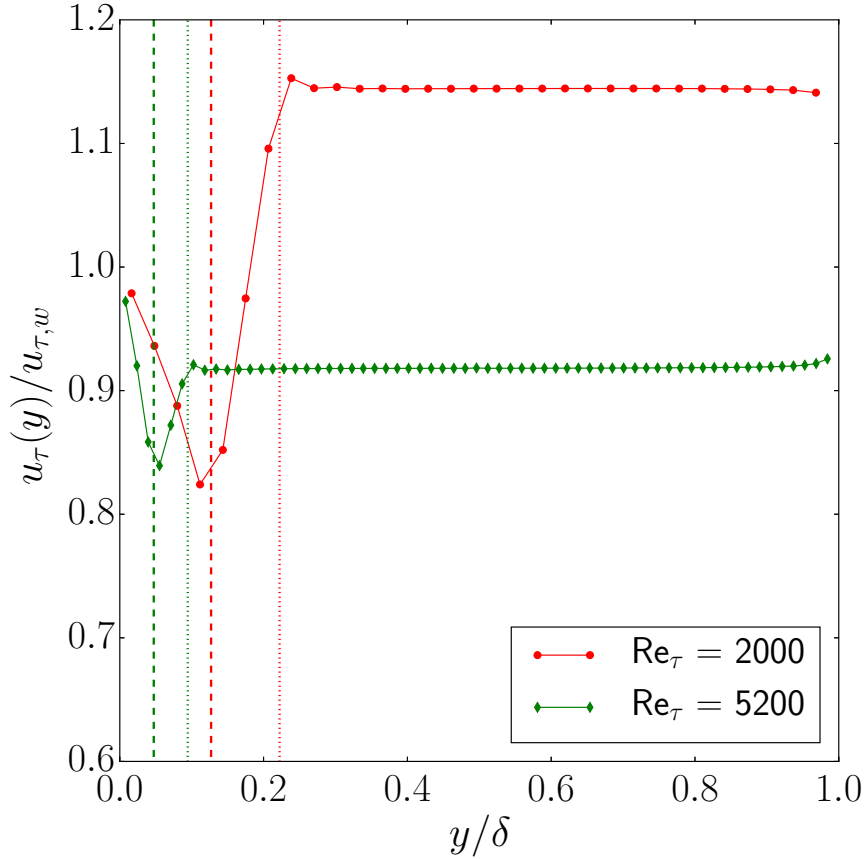}
	\caption{Attainment of a constant velocity scale in the LES region. Profiles of local friction velocity normalized by the friction velocity at the wall are shown. Both Re$_{\tau}$ = 2000 and Re$_{\tau}$ = 5200 are simulated with the tri-split forcing approach. Dashed vertical lines, $h_{RL}$; dotted vertical lines, $h_{sp}$.}
	\label{fig:LocalUtau}
\end{figure}

Tuerke and Jimenez\cite{tuerke2013simulations} proposed to use the ``local'' friction velocity to collapse Reynolds stress profiles in their inverse RANS (IRANS) study where mean velocity profiles were prescribed as a forcing term to a DNS. We follow the same approach and calculate the local friction velocity as follows:
\begin{align}
u_{\tau} (y) = \sqrt{\frac{\langle \tau (y) \rangle}{1 - y / \delta}}.
\label{eq:LocalFrictionVelocity}
\end{align}
The variation of the friction velocity along the wall normal distance is shown in Fig. \ref{fig:LocalUtau} for Re$_{\tau}$ 2000 and 5200.  A consistent feature in both friction velocity profiles is that a constant value emerges above the RANS-LES split height, $h_{sp}$. We refer to this constant value in the core LES region as $u_{\tau,c}$. 
\begin{figure}
	\centering
	\subfigure{ \includegraphics[width=0.48\textwidth]{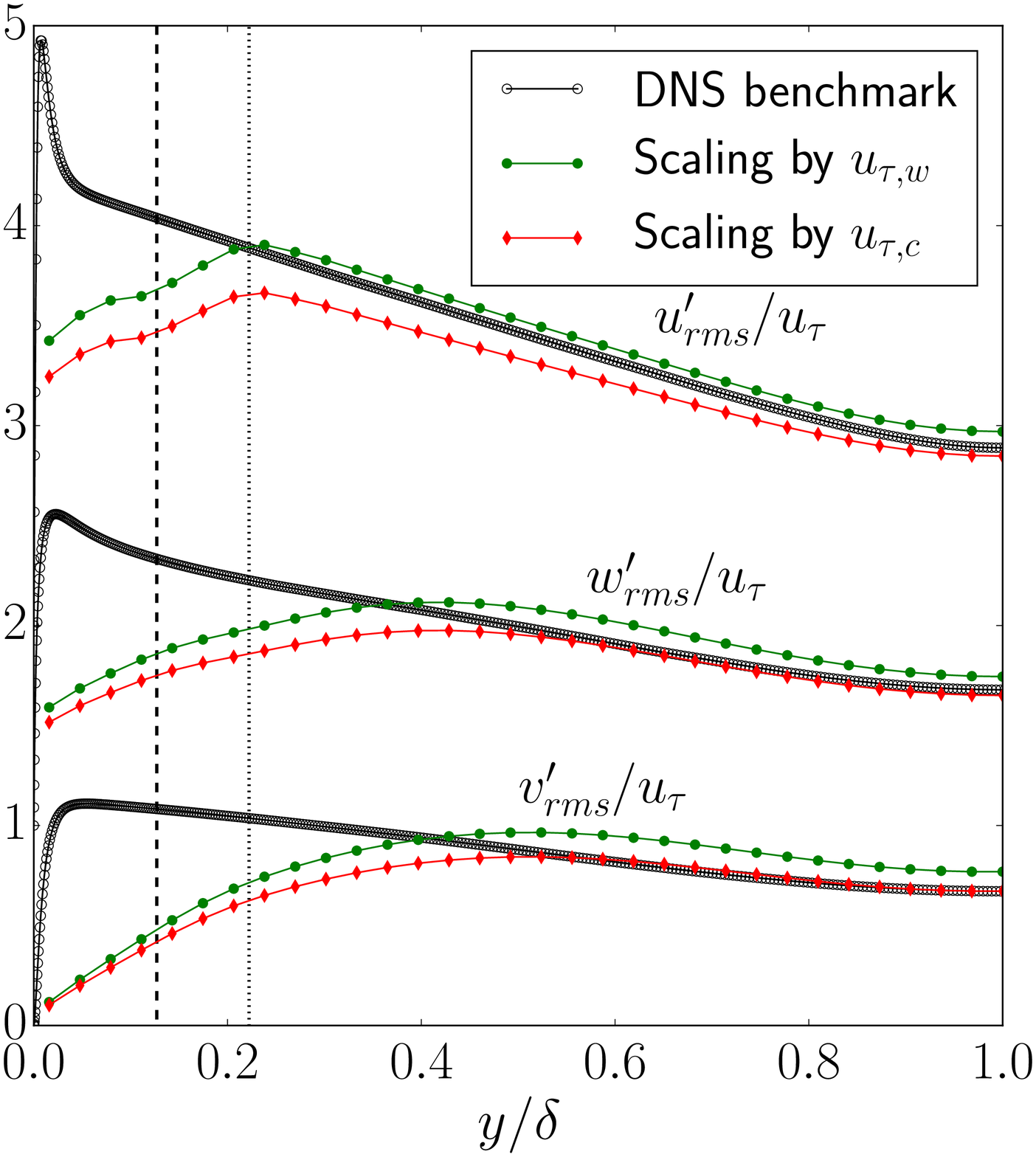}\label{fig:CoreUtauUU2000}}
	\subfigure{ \includegraphics[width=0.48\textwidth]{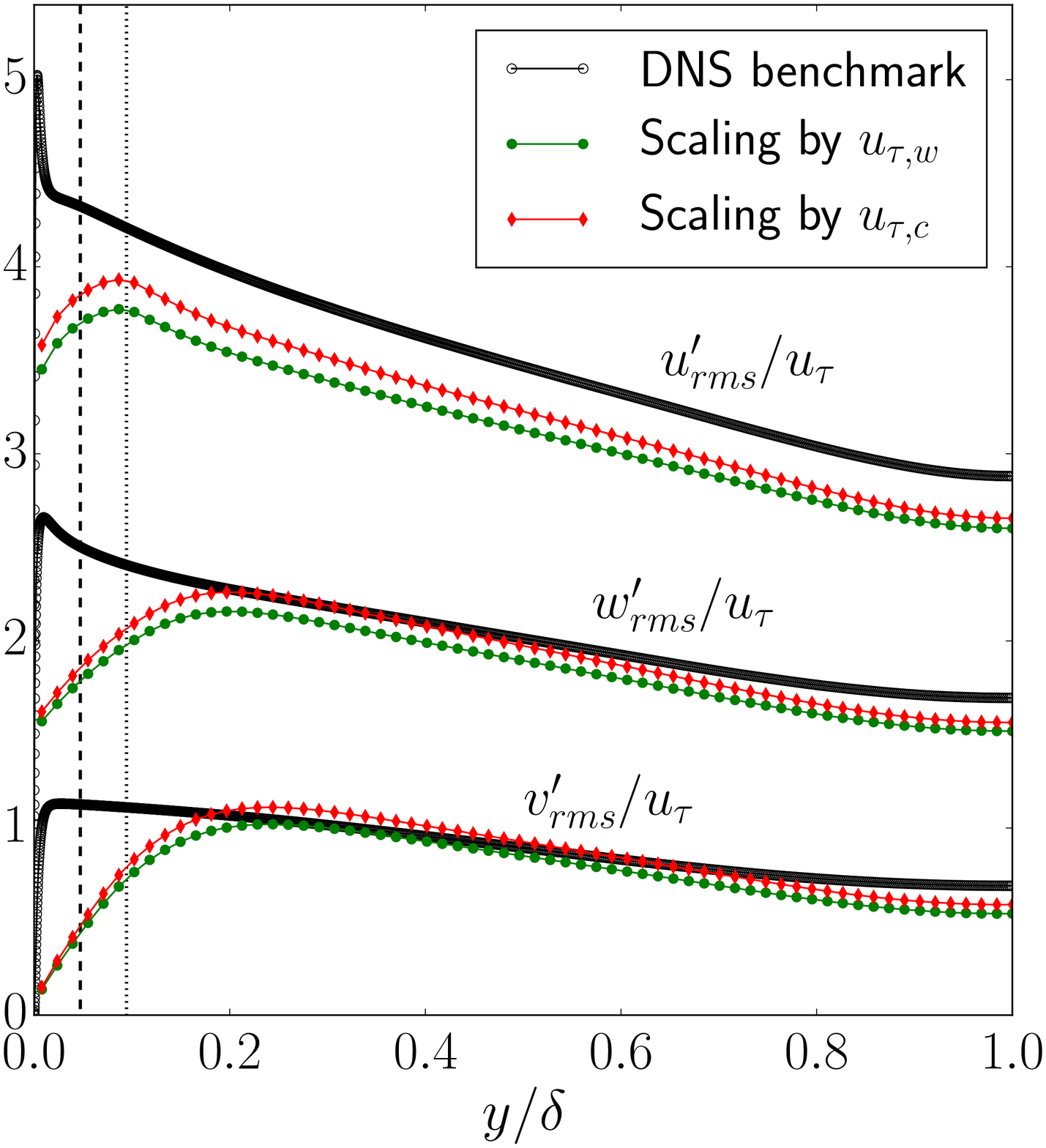}\label{fig:CoreUtauUU5200}}
	\caption{Profiles of rms velocity fluctuations and the effect of velocity scale choice on the results. (\textit{a}) Re$_{\tau}$ = 2000 and (\textit{b}) Re$_{\tau}$ = 5200. Streamwise and spanwise components shifted up by 2.0 and 1.0, respectively. Dashed vertical lines, $h_{RL}$; dotted vertical lines, $h_{sp}$.}
\end{figure}
We study the implications of a constant $u_{\tau,c}$ in the LES core by assessing the second-order statistics of turbulence. In Figs. \ref{fig:CoreUtauUU2000} and  \ref{fig:CoreUtauUU5200}, we compare normalized RMS velocity fluctuations with the scaling factors of $u_{\tau,c}$ and $u_{\tau,w}$. For Re$_{\tau}$ = 2000 in Fig. \ref{fig:CoreUtauUU2000}, the normalization by $u_{\tau,c}$ shows a clear improvement over scaling by $u_{\tau,w}$ for the spanwise and wall-normal components, matching the DNS benchmark data very well. The agreement with the streamwise component $u^{\prime}_{rms}$ is slightly worse. However, the $u^{\prime}_{rms}$ profile normalized by $u_{\tau,c}$ does not overestimate the DNS data, which makes sense as this is a filtered quantity obtained from a coarse mesh. Also for Re$_{\tau}$ =5200, $u^{\prime}_{rms}$ results scaled by $u_{\tau,c}$ are closer to the DNS normalized values than they are when scaled by $u_{\tau,w}$. Turbulent motions are damped by the large bulk eddy-viscosity as expected, therefore we do not seek any agreement with DNS data near the wall. Relative to the streamwise component, the spanwise and wall-normal components of the RMS velocity fluctuations take some distance past $h_{sp}$ before agreeing with the DNS benchmark data in the core region of the channel, suggesting that the three-dimensionality of turbulence emerges at some distance away from the RANS-LES interface. The lack of agreement below $y/\delta <$  0.4 for Re$_{\tau}$ = 2000 and $y/\delta <$ 0.2 for Re$_{\tau}$ = 5200 in second-order statistics is not unexpected as it is an outcome of adopting a Reynolds-averaging approach close to the wall. We note that our focus in this work to investigate the role of bulk eddy-viscosity variation on the log-layer mismatch problem and not to develop a new turbulence model.

\begin{figure}
	\centering
	\subfigure{\includegraphics[width=0.48\textwidth]{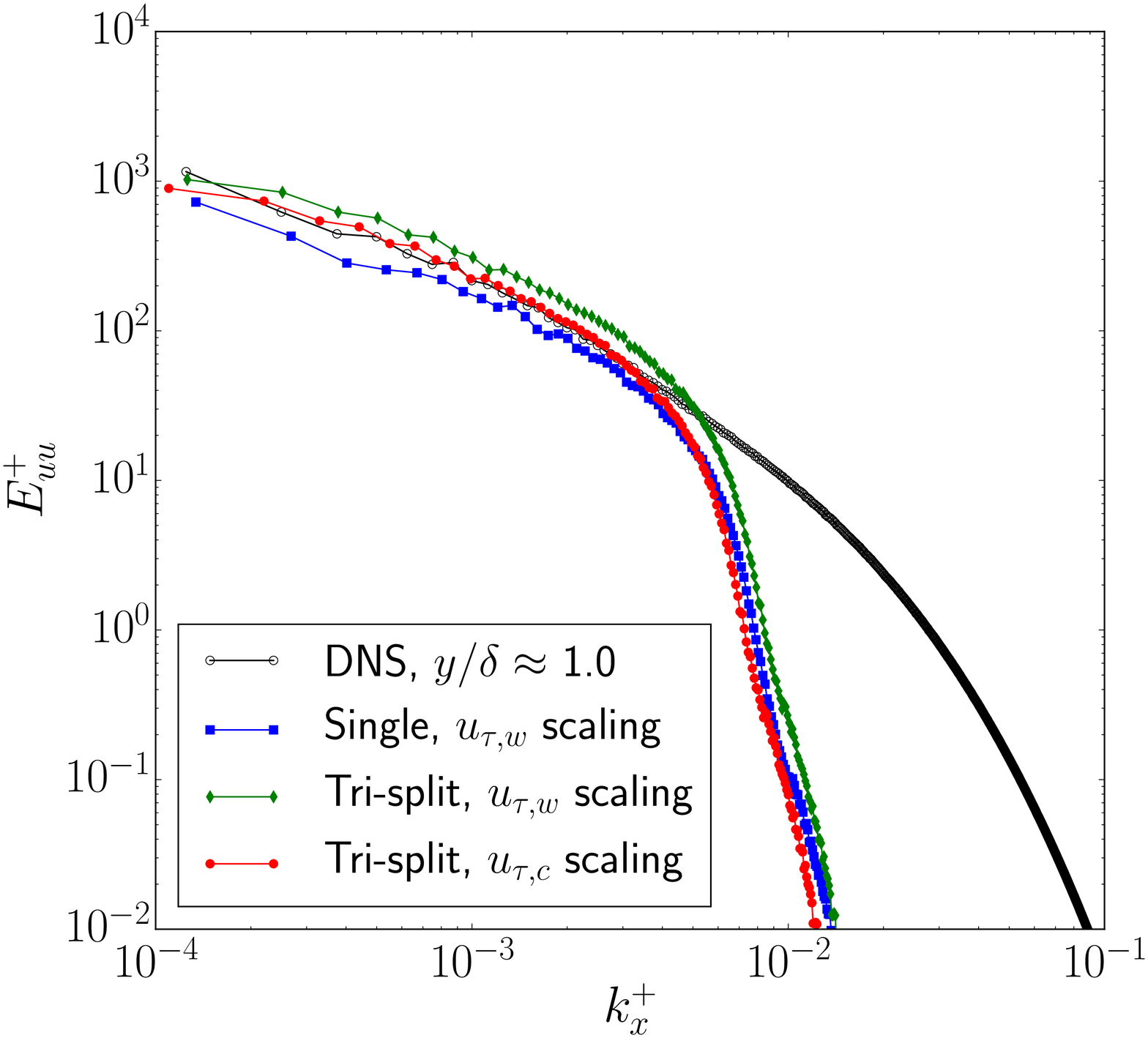}\label{fig:spectra2000}}
	\subfigure{\includegraphics[width=0.48\textwidth]{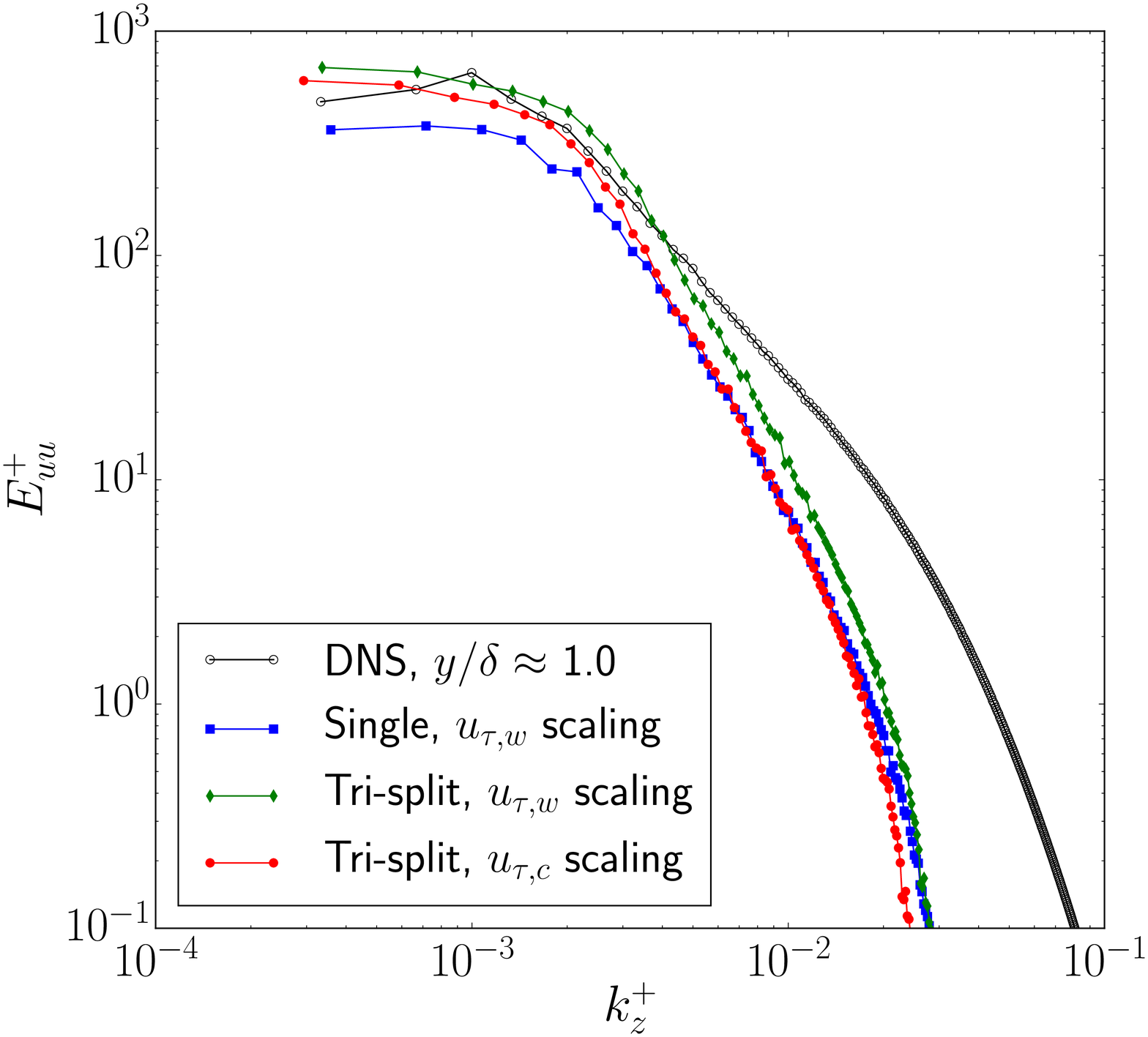}\label{fig:spectra5200}}
	\caption{Streamwise velocity spectra. Streamwise velocity spectra for Re$_{\tau}$ = 2000 (\textit{a}) in streamwise direction (\textit{b}) in spanwise direction.}
	\label{fig:spectra}
\end{figure}

One-dimensional normalized streamwise velocity spectra shown in Fig.\ref{fig:spectra} further supports our arguments about using  $u_{\tau,c}$ to scale turbulent fluctuations. Fig.\ref{fig:spectra} presents $E^{+}_{uu}$, for Re$_{\tau}$ = 2000 at the channel centerline for both streamwise and spanwise directions. The simulations capture well the large eddies as well as a portion of the inertial subrange at the lower wavenumbers. Note that it is not realistic to expect capturing the spectra at higher wave numbers because of the coarse mesh at hand. The simulation with tri-split forcing when scaled with $u_{\tau,c}$ agrees better with the normalized DNS spectra than both the single forcing and the tri-split forcing when scaled with $u_{\tau,w}$.

From the profiles of first-order and second-order statistics of turbulence, we observe two distinct velocity scales to nondimensionalize the results: $u_{\tau,w}$ and $u_{\tau,c}$. We emphasize that this duality in velocity scale is an artifact of the zonal enforcement of the constant mass flow-rate through the channel, and it does not come from the actual physics of turbulent channel flows. The only aspect that is consistent with expected flow physics is that $u_{\tau,c}$ attains a constant value in the LES core above the Reynolds-averaged zone. It then makes sense to use $u_{\tau,c}$ to scale the fluctuations in the LES region, as this region develops above a RANS region with its own forcing. But a worthy question comes up: Why is $u_{\tau,w}$ a better velocity scale for the mean velocity profile? To answer this question we further analyze the mean force balance.

\begin{table}
	\begin{center}
	\caption{Normalized values for verification of Eq. \ref{eq:intforcings}. Each subscript represents a forcing region. All forcing values are normalized by $u_{\tau,w}$ and $\delta$. All heights are normalized by $\delta$. Mean forcing values can be found in Figs. \ref{fig:ForcingTimeSeries2000} and \ref{fig:ForcingTimeSeries5200}. }
	\begin{tabular}{ c c c c c c c c }
		\hline 
		Re$_{\tau}$ & $h_1$ & $h_2$ & $h_3$ & $<f_1>$ & $<f_2> $ & $<f_3>$ & $\sum_{i=1}^{3} <f_i> h_i$ \\ 
		\hline
		2000 & 0.127 & 0.095 & 0.778 & 3.502 & -4.873 & 1.312 & 1.001 \\
		5200 & 0.047 & 0.047 & 0.906 & 7.496 & -2.445 & 0.841 & 1.000 \\
		\hline 
	\end{tabular}
	\label{tab:values}
	\end{center}
\end{table}

As the split regions evolve under the constraint of maintaining the target mass flow-rate, a global force balance is satisfied as follows
\begin{align}
\label{eq:intforcings}
\left\langle f_1 \right\rangle h_{1} + \left\langle f_2 \right\rangle h_{2} + \left\langle f_3 \right\rangle h_{3}) = \tau_{w} = \frac{u_{\tau,w}^2 }{\delta} \delta = f_x \delta,
\end{align}
where $h_1$ = $h_{RL}$, $h_2$ = $h_{sp} - h_{RL}$, and $h_3$ = $\delta - h_{sp}$. From the above relationship, we clearly see that $u_{\tau,w}$ is an average velocity scale for the entire channel flow when it is forced zonally. Table \ref{tab:values} presents the normalized values that can be used to verify the above equation. Therefore, it is not surprising that mean velocity profiles across the channel agree well with the DNS benchmark when normalized by the $u_{\tau,w}$. It is also evident why we cannot use $u_{\tau,c}$ to scale the entire velocity profile, because it does not satisfy the global force balance as written in Eq. \ref{eq:intforcings}.  

\begin{figure}
	\centering
	\subfigure{\includegraphics[width=0.48\textwidth]{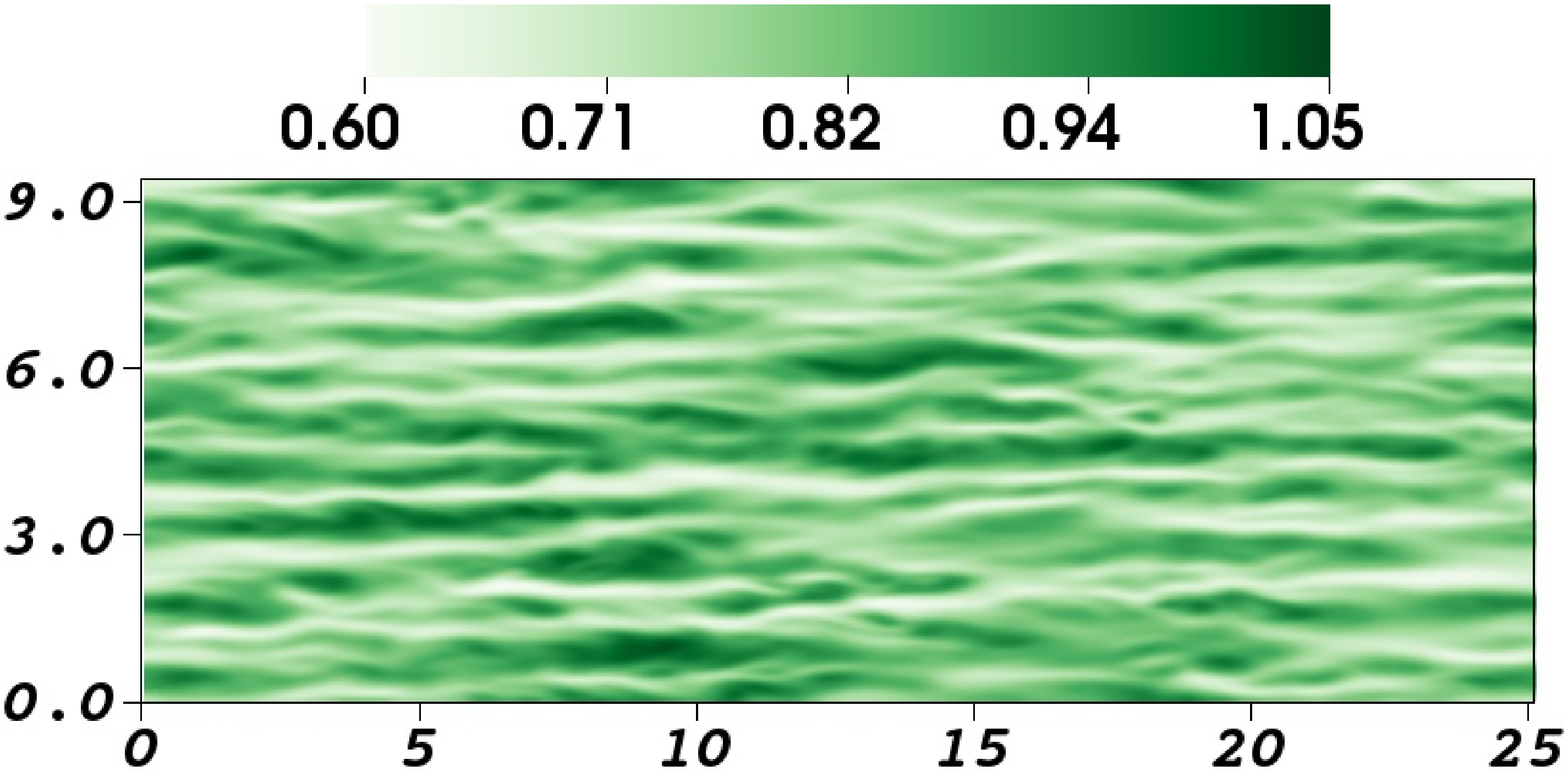}\label{fig:SingleStreaks_065}}
	\subfigure{\includegraphics[width=0.48\textwidth]{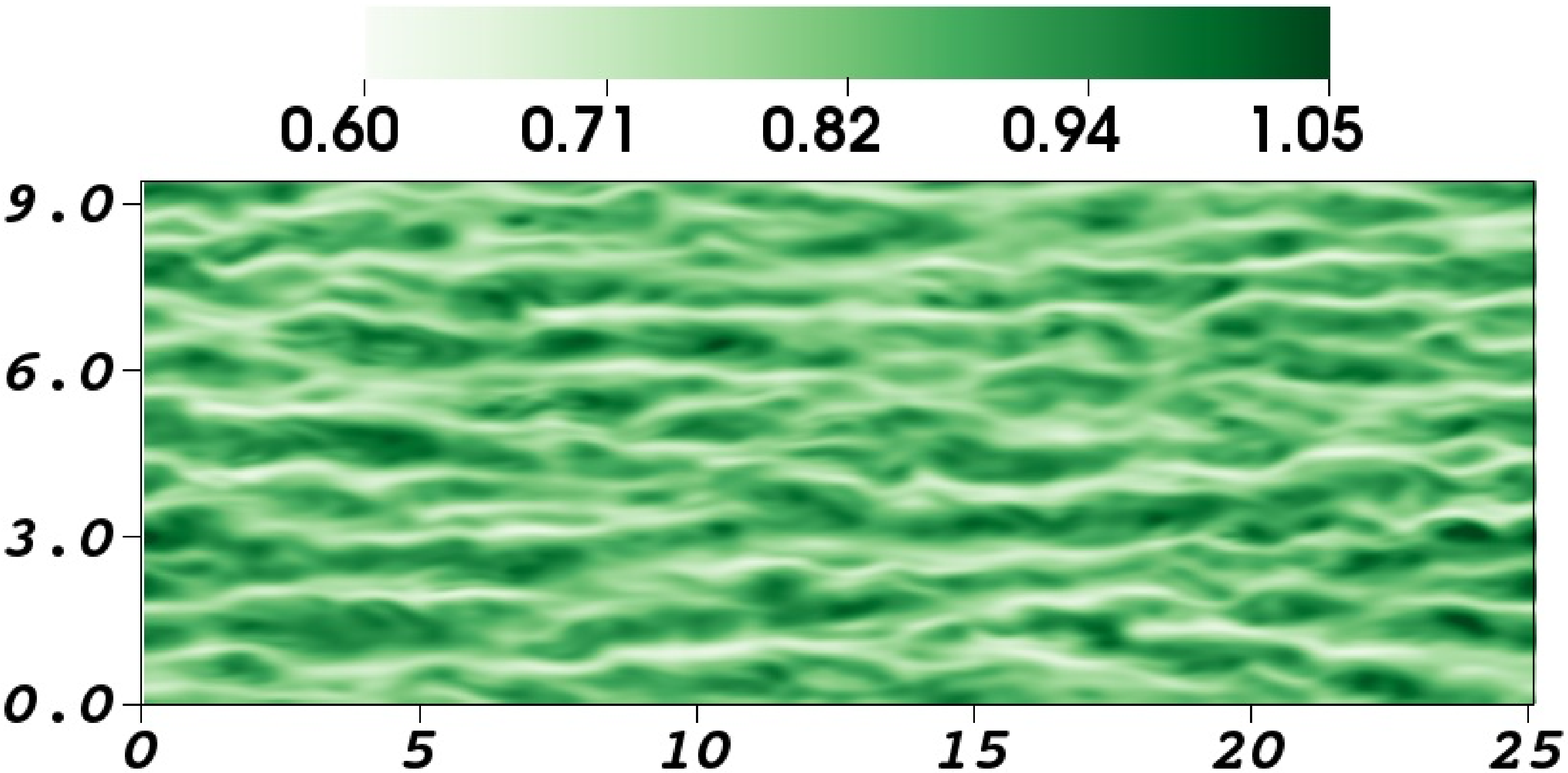}\label{fig:TripleStreaks_065}}
	\\
	\subfigure{\includegraphics[width=0.48\textwidth]{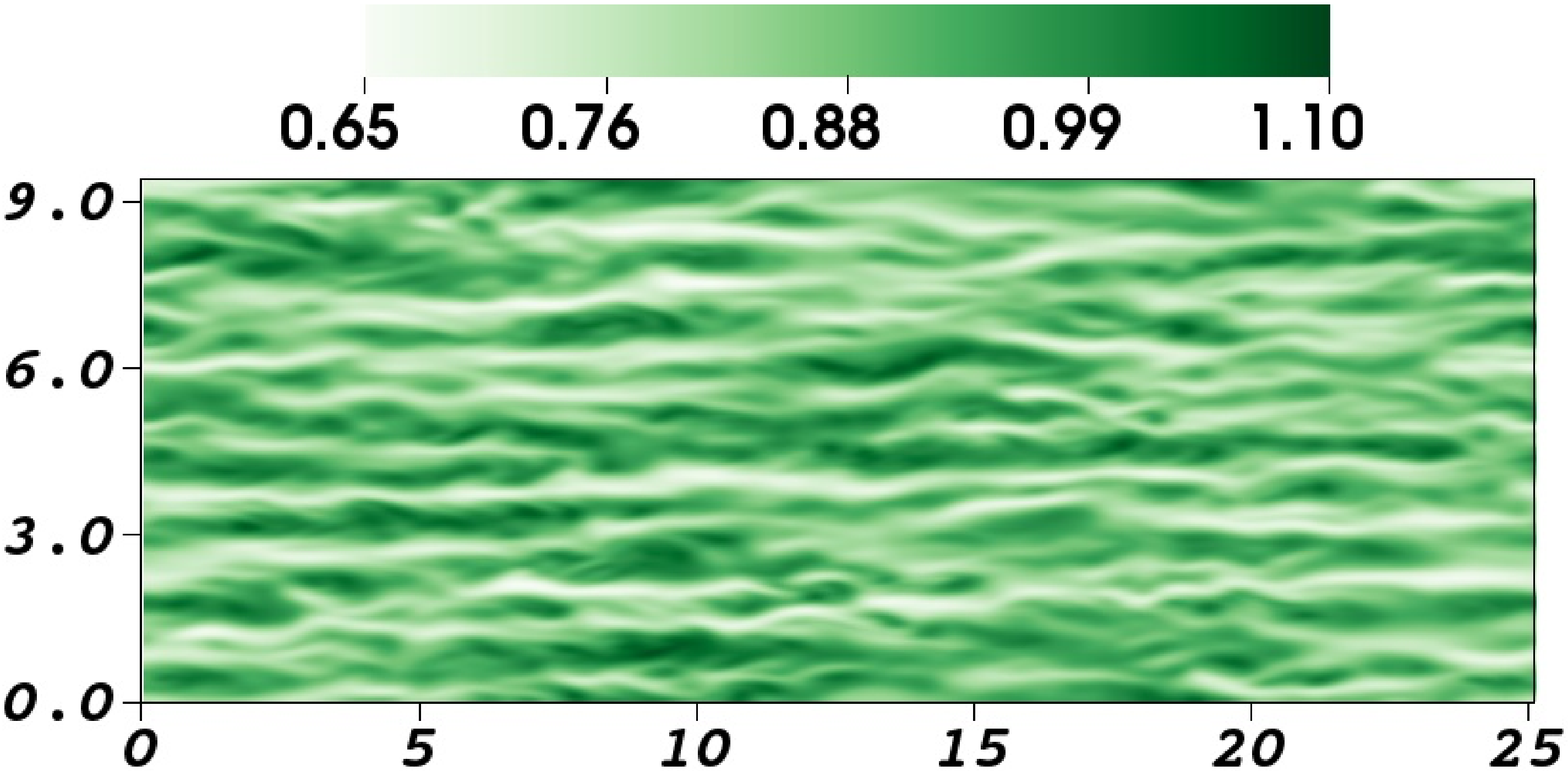}\label{fig:SingleStreaks_090}}
	\subfigure{\includegraphics[width=0.48\textwidth]{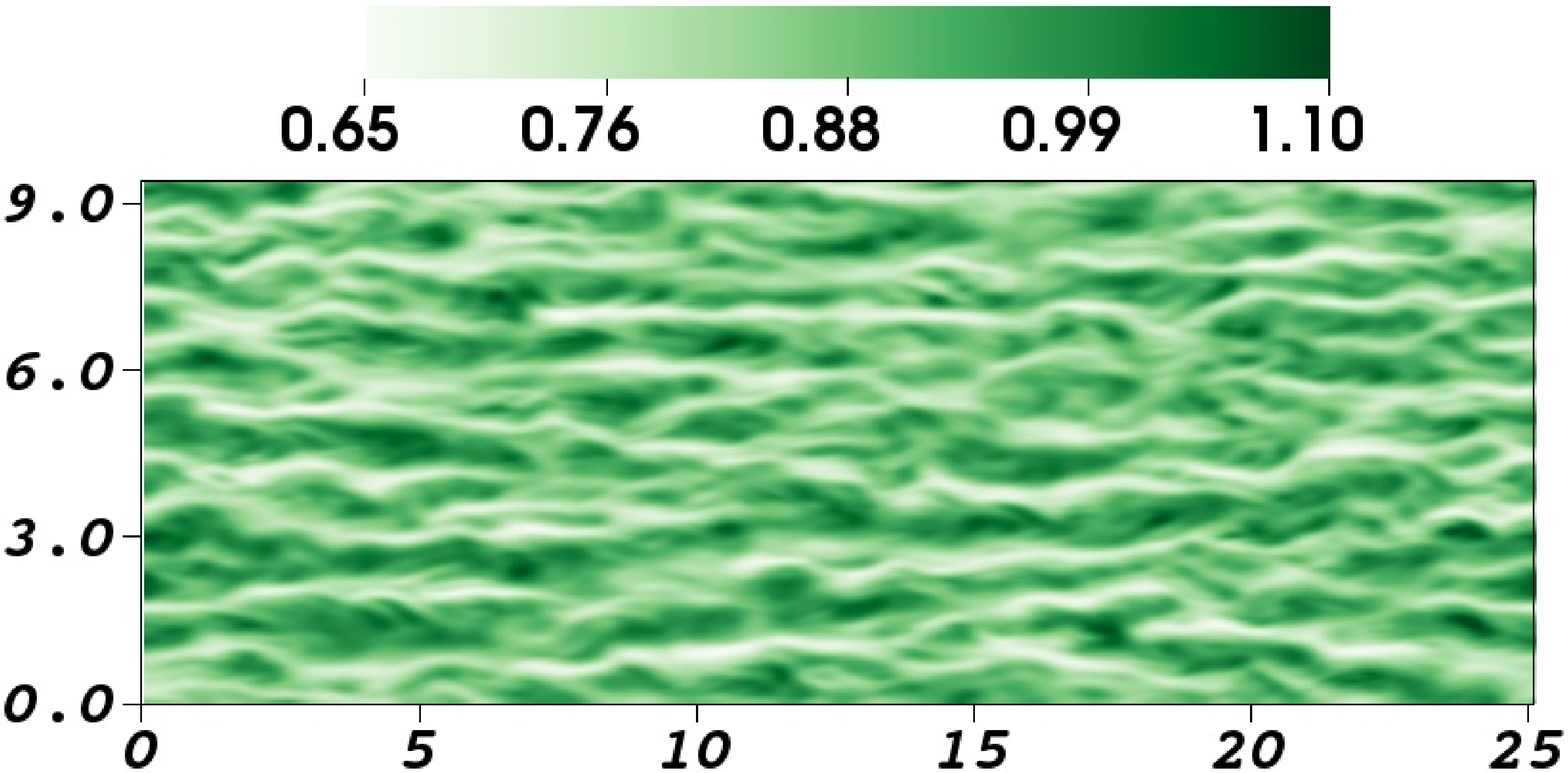}\label{fig:TripleStreaks_090}}
	\\
	\subfigure{\includegraphics[width=0.48\textwidth]{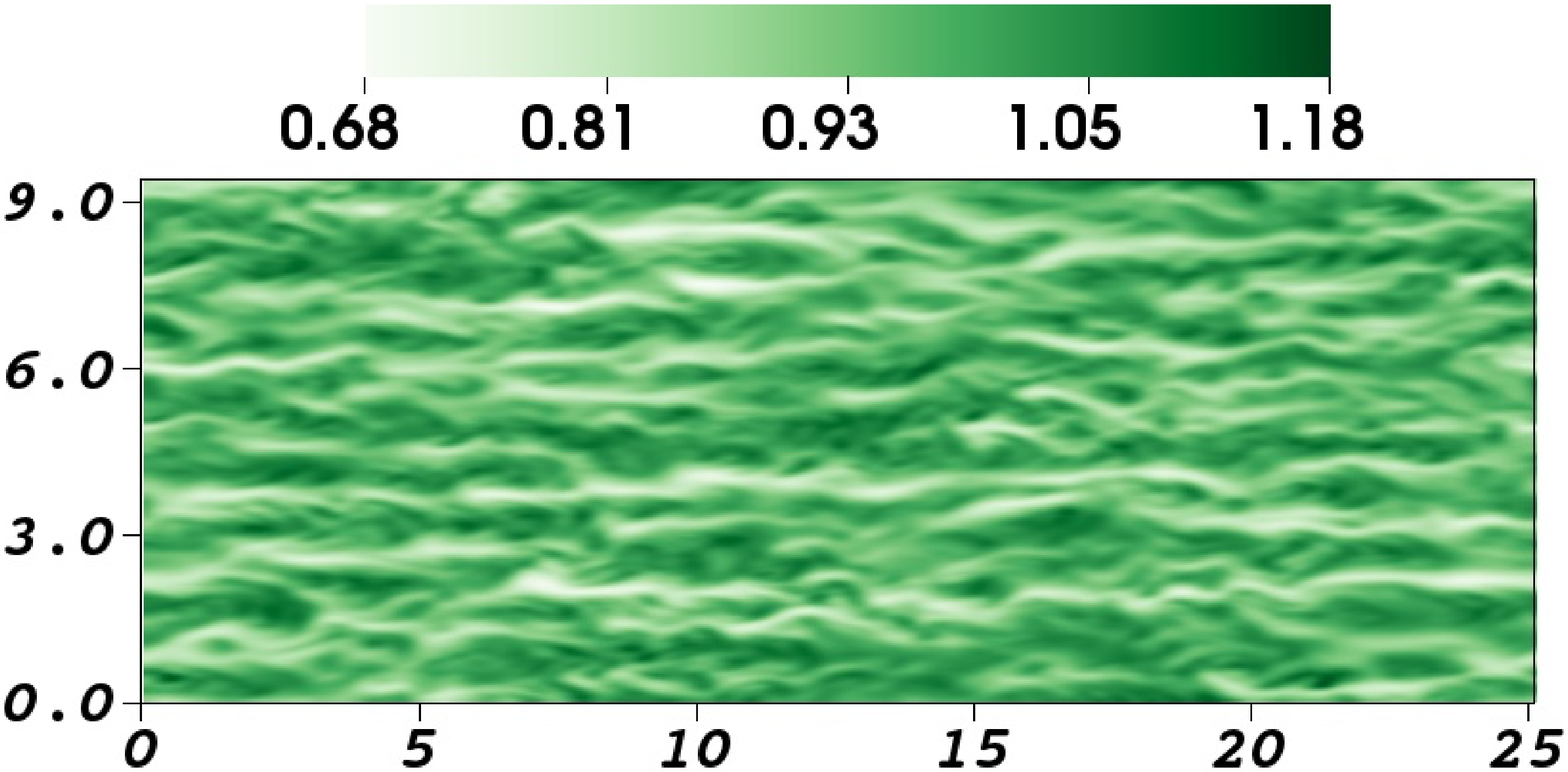}\label{fig:SingleStreaks_135}}
	\subfigure{\includegraphics[width=0.48\textwidth]{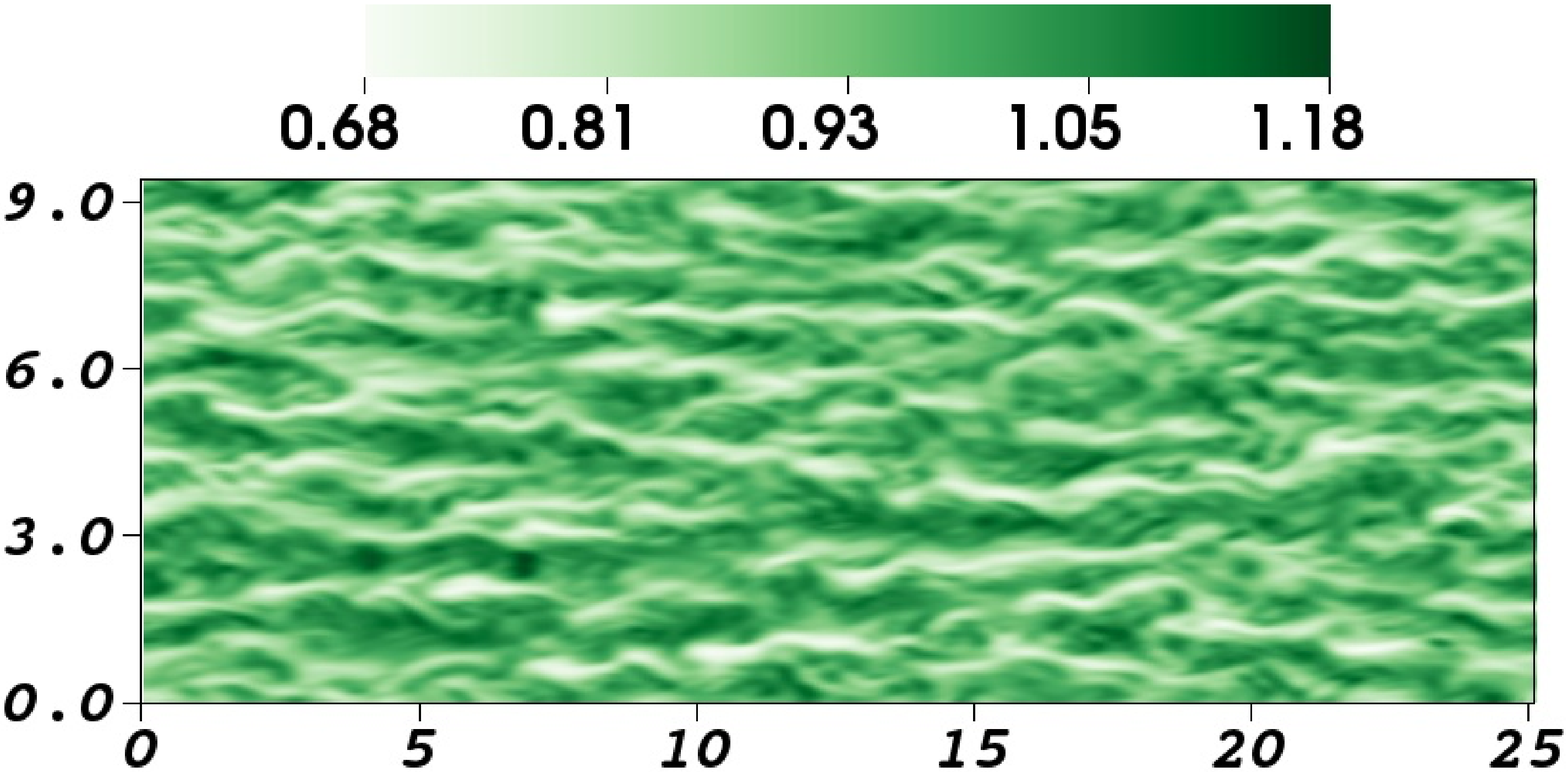}\label{fig:TripleStreaks_135}}
	\caption{Visualization of instantaneous streamwise velocity for Re$_{\tau}$ = 2000 at wall-normal heights of: (\textit{a}) and (\textit{b}) y / $\delta \approx$ 0.13; (\textit{c}) and (\textit{d}) y / $\delta \approx$ 0.18; (\textit{e}) and (\textit{f}) y / $\delta \approx$ 0.27. Re$_{\tau}$ = 2000. Single forcing results are shown in the left column and tri-split forcing results are shown in the right column. All velocities and lengths are normalized by the bulk velocity and channel half-height, respectively.}
	\label{fig:viz}
\end{figure}

The formation of the so-called ``super-streaks'' near the RANS-LES interface has been mentioned in earlier studies as a modeling artifact \cite{piomelli2003inner, keating2006dynamic}. We observe that the current split-forcing approach is able to break up these ``super-streaks'' as shown in Fig. \ref{fig:viz} where $y$ / $\delta \approx$ 0.13 - 0.18. The coarseness of the current mesh does not accommodate  formation of physical streaks, but the ability for the split-forcing to reduce the size of the unphysical streaks relative to single forcing simulation is encouraging. Away from the RANS-LES interface, both the tri-split forcing and the single forcing produced similar flow structures.

\section{Conclusions}

In this study, we showed that bulk eddy-viscosity variation can also contribute to log-layer mismatch by reducing the degrees of freedom in the mean momentum-balance equation. Unlike other studies where additional degrees of freedom were introduced through explicit stochastic terms, we imposed the target mass flow-rate through the channel zonally without resorting to an external forcing term. As a result, we have observed a significant reduction in the log-layer mismatch despite the coarseness of the computational mesh at hand. Unphysical streaks at the RANS-LES interface have diminished in size, and skin friction coefficients have improved markedly as well. 

The zonal enforcement of the mass flow-rate through the channel allowed a distinct velocity-scale to emerge in the LES core above the Reynolds-averaged region as a constant value. Both the second-order statistics and the one-dimensional velocity spectra agreed better with the DNS benchmark data when normalized by this constant velocity scale extracted from the LES core. The existence of a constant velocity scale supports the formation of an artificial boundary layer above the Reynolds-averaged region as originally suggested by \citet{baggett1998feasibility}.

The current approach to enforce the mass flow-rate zonally is limited to the turbulent channel flow problem, but it could be used to study the nature of turbulent fluctuations in the outer layer for very high Reynolds numbers without taxing computational resources. 

\section*{Acknowledgments}
This material is based upon work supported by the National Science Foundation under Grant No. 1056110 and 1229709. The first author would like to acknowledge the generous support he received through the University of Idaho President's Doctoral Scholars Award. We also like to thank Prof. Ralph Budwig for helpful discussions and his continuous mentorship.

%

\end{document}